\documentclass[twocolumn,amsmath,amssymb]{revtex4}
\usepackage{graphicx}
\begin{document}

\title{Domain wall Dynamics of the Ising chain in a Transverse field}
\author{V. Subrahmanyam}
\affiliation{
Max-Planck-Institut fuer Physik Komplexer Systeme,
Noethnitzer Str.38, Dresden 01187 Germany\\
and 
Department of Physics, Indian Institute of Technology, 
Kanpur 208016, India}
\altaffiliation[Permanent address, E-mail: vmani@iitk.ac.in\\]{}

\date{\today}

\begin{abstract} 
We show that the dynamics of an Ising spin chain in a transverse field
conserves the number of
domains (strings of down spins in an up-spin background) at discrete times.
This enables the
determination of the eigenfunctions of the time-evolution operator, and
the dynamics of initial states with domains. 
The transverse magnetization is shown to be identically zero 
in all sectors with a fixed number of domains. For an initial state with a 
single string of down spins,
the local magnetization, the equal-time and double-time spin-spin correlation 
functions,
are calculated analytically as functions of time and the initial string size. 
The domain size
distribution function can be expressed as a simple integral 
involving Bessel functions. 
\end{abstract}

\maketitle

The dynamics of the order parameter, and its correlation functions are
related to experimental observables of interacting systems, related to the 
relaxation and transport processes. For systems with known low-temperature
correlation functions, the dynamics can be extremely difficult to
determine. For example, the determination of correlators for
hard-core particles moving in one dimension has been done using the
inverse scattering method\cite{Korepin}.
The one-dimensional Ising model in a 
transverse field has a complicated dynamics, and real-time 
correlation functions have been shown to have a simple structure near a 
quantum critical point\cite{Young}. 

In this paper, we study the zero-temperature dynamics of a one-dimensional
transverse field Ising model. 
The dynamics does not conserve the spin
here, and generates domain walls, which can propagate and fluctuate in size.
The evolution of an initial state with one domain
of down spins (a string of successive down spins in the background of up spins),
involves creation of more domains and their complicated motion.
We show that the number of domains is conserved at discrete times. As a  direct
consequence the transverse magnetization is
identically zero for any state with a fixed number of domains. The calculation
of the one-point and two-point correlators can be carried out using the domain
eigenstates. In particular, the time evolution of initial states with exactly
one string of down spins is presented. 

Let us consider the Hamiltonian (on a ring of $N$ sites)
\begin{equation}
H\equiv H_z+H_x=-K\sum s_i^zs_{i+1}^z - \Delta \sum s_i^x+{NK \over 4}
\end{equation}
where $s_i^z,s_i^x$ are the components of the spin-1/2 operator at site $i$,
$K$ is the Ising interaction strength and $\Delta$ is the magnetic field 
strength
in x direction. A constant has been added for convenience, which we keep along
with the Ising interaction term in $H_z$. The above Hamiltonian can be 
diagonalized\cite
{Lieb}, and is extensively studied\cite{Pfeuty} (see \cite{Sachdev} for a 
recent review). Let us write the time evolution operator as (setting $\hbar=1$)
\begin{equation}
U=e^{-itH_z}e^{-i\int_0^t \tilde H_x(t^{\prime})}\equiv U_z\tilde U_x
\end{equation} 
where we have used the interaction representation, $\tilde H_x(t)=
\exp{(iH_zt)}H_x \exp{(-itH_z)}$. The time-evolved operator $\tilde s_i^x(t)$
can be expressed in terms of $s_i^x,s_{i+1}^z,s_{i-1}^z$, and we have
$$
\int_0^t \tilde s_i^x(t^\prime)dt^\prime=t\lbrack s_i^x (1-\hat a_i^2+{
\sin Kt\over Kt} \hat a_i^2) - s_i^y \hat a_i {\sin^2{Kt\over 2}\over
 Kt/2}\rbrack $$ where the operator $\hat a_i=(s_{i+1}^z+s_{i-1}^z)^2$.
Now, let us consider discrete times
$t_n=n\tau $, where $\tau=2\pi/K$ (for $K\sim 0.01 eV$, $\tau\sim 10^{-13}$).
This simplifies the time evolution operator, which becomes $ U=
\exp{-itH_z}\exp{-itH_1}$,
\begin{equation}
H_1=\Delta \sum s_i^x({1\over2}-2s_{i+1}^zs_{i+1}^z).
\end{equation}
For these discrete times, the evolution operator $U_z$ has no role to play,
it gives unity as an eigenvalue acting on any state with any number of down
spins. The operator $H_1$ can only act at sites with one up and one down spin
on either
side, in which case it will flip the spin.  The polarized state, either all up 
or all down spins has no
dynamics, as it is an eigenstate of $H_1$ with eigenvalue zero. $H_1$ operates
on a state with a string of $m$ down spins to generate states with either
$m+1$ or $m-1$ down-spin strings. In time, the size of the string can change
(from $m=1$, a lone down spin, to $m=N-1$, a lone up spin)
and the center of mass of the spring can move in either direction.  Similarly
the operator acting on a state with many strings of down spins, causes them
to move and fluctuate in size, but the number of strings will remain the same.

Let us consider the subspace of one-string states, $|l,m>$ where $m=1,N-1$ 
denotes
the length of the down-spin string starting from the site $l=1,N$. By
arranging the basis states in the order $|1,1>..|1,N-1>,..|N,1>..|N,N-1>$,
the Hamiltonian has a structure of $N\times N$ tri-diagonal matrix with
periodic boundary conditions. The diagonal positions are occupied by a 
$N-1\times N-1$ tri-diagonal matrix with open boundary conditions, and
upper (lower) diagonal positions are occupied by a matrix with nonzero elements
only on the lower (upper) diagonal positions. All the nonzero matrix elements
are equal to $\Delta/2$.
The eigenstates
of the Hamiltonian $H_1$ can be easily constructed as
\begin{equation}
|{k,q}\rangle=A\sum_{l,m} e^{ik(2l+m)}\sin(qm)|{l,m}\rangle,
\end{equation} where $A$ is a normalization constant, and the two momenta
$q=\pi l/N,l=1,N-1$ and $p=\pi n/N,
n=1,N$ label the one-string eigenstates. The phase corresponds to the uniform
motion of the center of mass of  a string of $m$ down spins. The eigenvalue
for this state is $\epsilon (k,p)=2\Delta\cos k \cos q$.

The eigenfunctions are the same as
the two-magnon eigenfunctions of the XY model, or the two-fermion tight-binding
eigenfunctions, labeled by two momenta $0\le q_1,q_2\le 2\pi$, 
$\psi_{ij}=A e^{ip(j+i)}\sin q(j-i)$, using $i=l,j=l+m$ in the 
above, and the relative momentum $q=q_1-q_2/2$, and the average momentum
$p=q_1+q_2/2$. 
In the two-magnon eigenfunction the labels $i,j$ stand for the two
down spins, whereas here they stand for the first sites of the down-spin
and up-spin domains in the domain eigenfunctions. Also, the range of $p$ here
is $0\le p\le 2\pi$, whereas for the domain eigenfunctions, we have $0\le k
\le \pi$. So the two ranges
$0\le p \le \pi$ and $\pi p\le 2\pi$ of the two-fermion eigenfunctions map onto
one range $0\le k\le \pi$ of the domain eigenfunctions, and each value of
$p$ corresponds to two eigenstates (in terms of the fermions, the
two states are $ c_{q_1\uparrow}^\dag c_{q_2\downarrow}^\dag|0\rangle,
c_{q_1\downarrow}^\dag
c_{q_2\uparrow}^\dag|0\rangle$).
Similarly an n-string eigenstate maps onto a 2n-magnon
eigenstate of XY model (or a 2n-fermion tight-binding problem, with n fermions
with spin down, and spin configuration being a Neel configuration). It is
easy to check that for any one-string eigenstate, the expectation value
$<\vec s_i(t)>=0,$ for all times. And similarly for eigenstates in other 
sectors. However, the dynamics of an initial state with a given spin
configuration is quite complex. As we shall see below the transverse 
magnetization is still
zero, but, however, $\langle s_i^z\rangle$ can be nonzero.

Let us consider at $t=0$ a state with a string of $r_0$ down spins, starting
from a site $i_0$ and ending at the site $j_0=i_0+r_0$. At later
discrete times, the state can be written as
\begin{equation}
|\phi(t)\rangle=\sum \tilde\phi_{ij}(t)|i,j\rangle,
\end{equation}
where the wave function can be written in terms of the single-string
eigenfunctions. Let us write the two-fermion wave function as 
\begin{equation}
\phi_{ij}=\sum_{k,p}e^{-i\epsilon(k,p)t}\psi_{i_0,j_0} \psi_{ij}^\star,
\end{equation}
which becomes (after taking $N\rightarrow \infty$)
\begin{equation}
\phi_{ij}= e^{i{\pi\over 2}(i+j-i_0-j_0)}(J_{i-i_0}(x)J_{j-j_0}-J_{i-j_0}J_{j-i_0})
\end{equation} where the argument of all the Bessel functions above is
$x=2\Delta t$. And $\tilde\phi_{ij}=\phi_{ij}$ for $j>i$, whereas
$\tilde\phi_{ij}=\phi_{i,N+j}$ for $i>j$ (here the length of the down-spin
string is $N-i+j$). Though the fermion wave function is the same for both
cases, but for the domains it is quite different, $j<i$ configurations occur
with negligibly small probabilities for a large $N$. The expectation values of
operators at
time $t$ reduce to sums over Bessel functions. Let us first consider the
transverse magnetization, $<s_i^x(t)>=<s_i^++s_i^->/2$. The raising and
lowering
operators connect basis states that differ by a unit string length, and we have
\begin{equation}
<s_i^->=\sum \phi_{i,m}^{\star} \phi_{i+1,m} +\phi_{i+1,m}^{\star}\phi_{i,m}=0,
\end{equation} and similarly the raising operator has zero expectation value.
Both the components of transverse magnetization vanish identically here.
For the case of states with more number of strings, the argument is as follows.
As the wave functions are 2n-fermion wave functions, a nonzero expectation
value for $<s_i^->$ would imply an off-diagonal long-ranged order; which is
prohibited by an argument due to Yang\cite{Yang}, viz. the off-diagonal 
elements of a one-particle reduced density matrix cannot exhibit long range
order. For a nonvanishing transverse magnetization, one has to construct
superpositions of states drawn from different domain-number sectors. These
states will be more meaningful for the low-temperature behaviour for 
$\Delta>K$.
\begin{figure}
\begin{picture}(0,0)%
\includegraphics{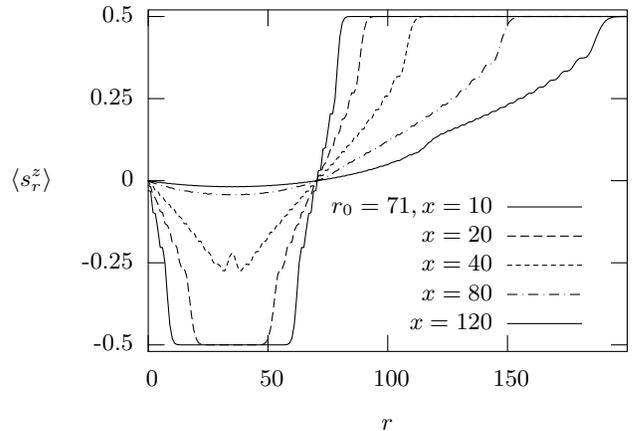}%
\end{picture}%
\setlength{\unitlength}{0.0200bp}%
\begin{picture}(12600,8640)(0,0)%
\put(2475,1774){\makebox(0,0)[r]{\strut{}-0.5}}%
\put(2475,3322){\makebox(0,0)[r]{\strut{}-0.25}}%
\put(2475,4870){\makebox(0,0)[r]{\strut{} 0}}%
\put(2475,6418){\makebox(0,0)[r]{\strut{} 0.25}}%
\put(2475,7966){\makebox(0,0)[r]{\strut{} 0.5}}%
\put(2750,1100){\makebox(0,0){\strut{} 0}}%
\put(5006,1100){\makebox(0,0){\strut{} 50}}%
\put(7263,1100){\makebox(0,0){\strut{} 100}}%
\put(9519,1100){\makebox(0,0){\strut{} 150}}%
\put(550,4870){\rotatebox{0}{\makebox(0,0){\strut{}$\langle s_r^z\rangle$}}}%
\put(7262,275){\makebox(0,0){\strut{}$r$}}%
\put(9244,4375){\makebox(0,0)[r]{\strut{}$r_0=71,x=10$}}%
\put(9244,3825){\makebox(0,0)[r]{\strut{}$x=20$}}%
\put(9244,3275){\makebox(0,0)[r]{\strut{}$x=40$}}%
\put(9244,2725){\makebox(0,0)[r]{\strut{}$x=80$}}%
\put(9244,2175){\makebox(0,0)[r]{\strut{}$x=120$}}%
\end{picture}%
\caption{The local magnetization as a function of the position $r=i-i_0$ for
an initial state with $r_0=71$ down spins, for various values of time $
x=2 \Delta t$.}
\end{figure}
 
Let us turn to the magnetization at a given site, and since $s_i^z$ is 
diagonal in this basis, we can express its expectation value as, 
\begin{equation}
\langle s_i(t)\rangle = {1\over 2}(\sum^{i-1}\sum_{l+1}^i +\sum_{i+1}\sum_{l+1}
^{N} -\sum^i\sum_{i+1}^N)|\phi_{lm}|^2.
\end{equation} Let us denote $i=i_0+r,j_0=i_0+r_0$, and choose $i_0=N/2$ for
convenience. Since $r=-a, r=r_0-1+a$ are equidistant locations from the 
string on either side of the string, those two sites will
have the same magnetization for all times. Thus, it suffices to study
$r>0$. We have,
\begin{equation}
\langle s_r^z \rangle ={1\over2} (f_1 f_2 -f^2)
\end{equation} where $f_1=g(r), f_2=g(r-r_0)$, for $r\ge r_0,
f_2=-g(r_0-r-1),$ for $0<r<r_0$, and the function $g(r)=\sum_{-r}^{r} J_n^2(x)$.
It can be expressed as
\begin{equation}
g(r)=1-2\int_0^x J_r(x_1)J_{r+1}(x_1)dx_1.
\end{equation}The function $f(r,r_0)=2\sum_{r+1}^\infty J_nJ_{n-r_0}$ can
be written as
\begin{eqnarray}
f(r,r_0)&&=\int_0^x (J_rJ_{r+1-r_0}+J_{r+1}J_{
r-r_0}) dx_1 \nonumber \\
&&={x\over r_0}( J_rJ_{r+1-r_0}-J_{r+1}J_{r-r_0}(x) )
\end{eqnarray} For a fixed value of $x$ as $r\rightarrow\infty$,
$g\rightarrow 1,f\rightarrow 0$, which implies $<s_r^z>\rightarrow 1/2$. This
is because the domain wall will take $x\sim r_0+r$ time to reach the far away
point. For fixed $r$, as $x\rightarrow\infty,$ the two functions have the
behaviour $g\rightarrow 0,f\rightarrow 2/\pi r_0$, for $r_0$ odd, and it
is zero when $r_0$ is even.  This implies all the spins
within a size of $x$ from the initial position have an expectation value 
$-2/\pi^2r_0^2$ for $r_0$ odd, whereas for even initial string sizes, the
local magnetization fluctuates
around a zero expectation value. The local magnetization is plotted as a 
function of the site index $r$, for various values of time in Fig.1, for
an intial state with a string of 71 down spins.

\begin{figure}
\begin{picture}(0,0)%
\includegraphics{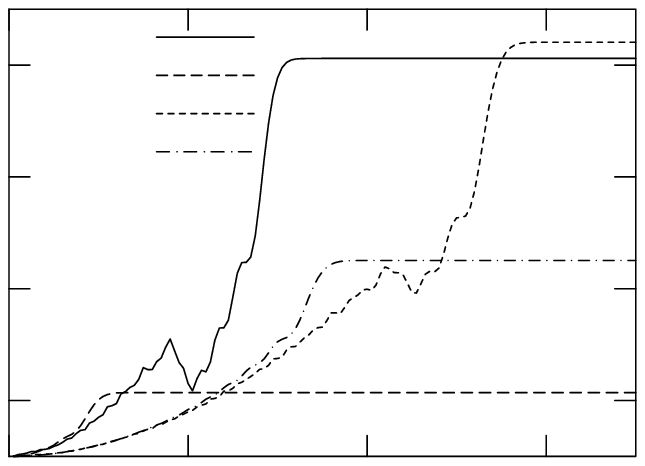}%
\end{picture}%
\setlength{\unitlength}{0.0200bp}%
\begin{picture}(12600,8640)(0,0)%
\put(2475,2455){\makebox(0,0)[r]{\strut{} 0.02}}%
\put(2475,4065){\makebox(0,0)[r]{\strut{} 0.06}}%
\put(2475,5675){\makebox(0,0)[r]{\strut{} 0.1}}%
\put(2475,7285){\makebox(0,0)[r]{\strut{} 0.14}}%
\put(2750,1100){\makebox(0,0){\strut{} 0}}%
\put(5329,1100){\makebox(0,0){\strut{} 40}}%
\put(7907,1100){\makebox(0,0){\strut{} 80}}%
\put(10486,1100){\makebox(0,0){\strut{} 120}}%
\put(550,4870){\rotatebox{0}{\makebox(0,0){\strut{}$\langle s_{r}^z s_{r+s}^z\rangle$}}}%
\put(7262,275){\makebox(0,0){\strut{}$s$}}%
\put(4602,7687){\makebox(0,0)[r]{\strut{}10,50}}%
\put(4602,7137){\makebox(0,0)[r]{\strut{}30,50}}%
\put(4602,6587){\makebox(0,0)[r]{\strut{}10,100}}%
\put(4602,6037){\makebox(0,0)[r]{\strut{}30,100}}%
\end{picture}%
\caption{The diagonal equal-time correlation function is plotted as a 
function of the
separation, for a few values of $(r,x)$. The initial domain size is
$r_0=30$ in all the cases.}
\end{figure}
The even-odd effect can be understood as follows. First we note that for $r_0$
odd (even), at later times odd (even) sizes are more probable than even (odd).
Later we shall compute the size distribution function.
The down-spin density is maximum at the center of the domain, and as we move
away from the center the spin density drops. Most of the contribution for 
a negative magnetization should be coming from the probability of a site being
a center of a domain. The center of mass of even-size
domains cannot be a lattice site. Hence, for even $r_0$, the magnetization
fluctuates around a zero mean. However, this
even-odd discrepancy disappears for larger initial string sizes, viz. for
$r_0=9$ the mean drops to $-0.002$ already. This is because as $r_0$ becomes
large, smaller-sized domain are less probable, as we shall see later. And
large odd-size domain move slower, thus depleting the probability for a site
to be a center of a domain. 

Now we turn to the calculation of equal-time correlation functions. The 
off-diagonal correlation function is straightforwardly written as $\langle
s_i^+s_j^-\rangle =\phi_{i+1,j+1}^\star \phi_{ij}$. 
The diagonal correlation function can be written as
\begin{equation}
\langle s_i^z(t)s_j^z(t)\rangle ={1\over 4}-{1\over2}(\sum_{i+1}^j\sum_{j+1}^N+
\sum_1^i\sum_{i+1}^j)|\phi_{lm}|^2.
\end{equation}Let us
choose, $i=i_0+r,j=i+s,j_0=i_0+r_0,$ and $i_0=N/2$ as before. Here there
are three regimes, depending whether both $i,j$ are within the initial string,
or both outside, or one of them is outside. The correlation function can be
reexpressed as
\begin{equation}
\langle s_r^z s_{r+s}^z\rangle={1\over 4}(F_1F_2-F^2)
\end{equation}where $F_1=g(r+s)-g(r)$ and $F=f(r,r_0)-f(r+s,r_0)$. The 
function $F_2$ has different dependencies in the three regimes, as,
$F_2=g(r-r_0+s)-g(r-r_0),$ for $r\ge r_0, F_2=g(r_0-r-1)-g(r_0-r-s-1),$ for
$ r+s<r_0, F_2=g(r_0-r-1)+g(r+s-r_0),$ for $r<r_0,r+s\ge r_0$.
For a fixed value
of $x$, as the separation $s\rightarrow \infty$, these functions reduce as 
$F_1\rightarrow f_1,F_2\rightarrow f_2,F\rightarrow f$.
And the two-point correlation function decouples, and becomes $\langle
s_r^zs_{r+s}
\rangle \rightarrow {1\over2}\langle s_r^z\rangle$ (for a fixed time, the far
away spin is up, as the domain wall has not had time to propagate till that
far). In Fig.2 the correlation function has been plotted as a function of
the separation, $s$, for a few values of $r$ and time, for initial states
with one domain of $r_0=30$ down spins. 
  
The double-time correlation functions are not very easy to evaluate. Let us
define $C_{ij}(t_2,t_1)=\langle \phi_0 |s_i^z(t_2)s_j^z(t_1)|\phi_0\rangle$. 
Setting $t_1=0,t_2=t$, we have
\begin{equation}
C_{ij}(t,0)=\langle s_i^z(t)\rangle \langle s_j^z(0)\rangle.
\end{equation}
The operator $s_j^z$ can be taken out, as $|\phi_0>$ is its 
eigenstate, and the rest is just $\langle s_i^z(t)\rangle$. But for
an arbitrary $t_1\ne 0$, the correlation function can be written as
\begin{eqnarray}
C_{ij}(t_2,t_1)&&=\sum_{l,m,l^\prime m^\prime} |\phi_{lm}(x_1)|\phi_{l^\prime
m^\prime}(x_2)|
\langle s_i^z\rangle_{l^\prime m^\prime}
\langle s_j^z\rangle_{lm}\nonumber \\
&&(J_{l^\prime-l}(x_2-x_1)J_{m^\prime-m}- J_{l^\prime-m}J_{m^\prime-l}).
\end{eqnarray}
In the above, the argument of all the Bessel functions is the difference
of time. The double-time correlation function is substantially more
complicated, and a detailed analysis will be presented elsewhere. 

\begin{figure}
\begin{picture}(0,0)%
\includegraphics{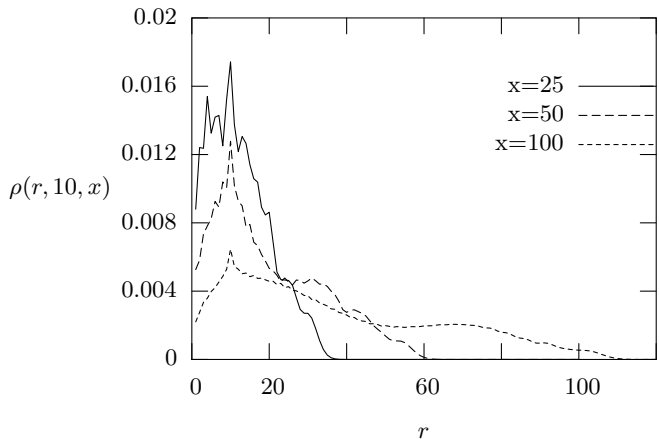}%
\end{picture}%
\setlength{\unitlength}{0.0200bp}%
\begin{picture}(12600,8640)(0,0)%
\put(2750,1650){\makebox(0,0)[r]{\strut{} 0}}%
\put(2750,2938){\makebox(0,0)[r]{\strut{} 0.004}}%
\put(2750,4226){\makebox(0,0)[r]{\strut{} 0.008}}%
\put(2750,5514){\makebox(0,0)[r]{\strut{} 0.012}}%
\put(2750,6802){\makebox(0,0)[r]{\strut{} 0.016}}%
\put(2750,8090){\makebox(0,0)[r]{\strut{} 0.02}}%
\put(3025,1100){\makebox(0,0){\strut{} 0}}%
\put(4483,1100){\makebox(0,0){\strut{} 20}}%
\put(7400,1100){\makebox(0,0){\strut{} 60}}%
\put(10317,1100){\makebox(0,0){\strut{} 100}}%
\put(550,4870){\rotatebox{0}{\makebox(0,0){\strut{}$\rho(r,10,x)$}}}%
\put(7400,275){\makebox(0,0){\strut{}$r$}}%
\put(10042,6802){\makebox(0,0)[r]{\strut{}x=25}}%
\put(10042,6252){\makebox(0,0)[r]{\strut{}x=50}}%
\put(10042,5702){\makebox(0,0)[r]{\strut{}x=100}}%
\end{picture}%
\caption{The size distribution function is plotted as function of the domain
size $r$, for a few values of time (the initial size is $r_0=10$ in all
cases).}
\end{figure}

We now turn to the domain-size distribution. The probability for a size 
$r$ to occur will depend on the initial domain size, and the time. We can
write the probability as $\rho(r,r_0,x)=\sum |\phi_{ij}|^2 \delta_{j,i+r}$,
which becomes, after replacing the delta function by a Fourier integral,
\begin{equation}
\rho={1\over \pi}\int_{-\pi}^{\pi} e^{i\phi r}\lbrack\cos{(\phi r_0)}h^2(0,x)
-h^2(r_0,x)\rbrack.
\end{equation} We have defined another auxiliary function involving a sum
over Bessel functions,
\begin{equation}
h(y,x)=|\sum_{-\infty}^{\infty}J_k(x) J_{k+y}(x)e^{ik\phi}|=
J_y(2x|\sin {\phi\over2}|),
\end{equation}
where we have used the summation rule for the Bessel functions in the last
step\cite{Gradshteyn}. After
a few more manipulations, we get
\begin{equation}
\rho={2\over \pi}\int_0^{1}{dt\over \sqrt{1-t^2}}\lbrack J_{r+r_0}(2xt) - (-1)^{r_0} 
J_{r-r_0}(2xt )\rbrack^2.
\end{equation} 
In Fig.3 we show the distribution function plotted against the
domain size for a few values of $x$, for an initial state with one
domain of down spins of size $r_0=10$. 
The probability is maximum for $r=r_0$, and as
$x$ becomes very large all sizes are equally probable modulo an even-odd
effect. The above distribution 
becomes simpler for the case of $r_0=1$ as
\begin{equation}
\rho(r,1,x)={8r^2\over \pi x^2}\int_0^{\pi/2} J_r^2(2x \cos \phi).
\end{equation} 
This distribution function can change
substantially for initial states with more than one string of down spins.
For a finite density $n$ of strings, the domain size $r$ is limited by the
average domain spacing in the initial state. The probability is more
for domain sizes $r<r_c\sim 1/n$. 

In conclusion, we have explored the domain-wall dynamics for the Ising model
in a transverse field in one dimension. The eigenstates of the time evolution
operator are
simple domain eigenstates for discrete times (multiples of $2\pi/K$). 
The transverse magnetization is identically zero in all sectors with different
number of domains. Explicit calculations of the local magnetization and the
spin-spin correlation functions are presented for initial states with one
domain of down spins. Though the above analysis is valid for large $\Delta$
also, these domain states do not belong to the low-energy spectrum here.
One can do a similar analysis with domain states using
the eigenstates of $s_i^x$ operators, which are more natural states for
$\Delta>K$.
Finally we would like to comment on the domains in two dimensions, where the
situation is quite different.
Firstly, the domains
can have various shapes. Secondly, though at discrete times, the time 
evolution can be
simplified analogously, however, the number of domains is not conserved.
Another major difference is that the domain size is limited, and they
do not move easily. For a domain
of arbitrary shape at some time, the domain size at later times is limited
by the size of the smallest rectangle into which the initial shape can be 
enclosed. 

It is a pleasure to thank Professor Peter Fulde for a critical reading of
the manuscript, and Professor Shen Shun-Qing for useful discussions.
  

\end{document}